\documentclass{llncs}
\usepackage{makeidx}  % allows for indexgeneration
\usepackage{pgf}  
\usepackage{xspace}  
\usepackage{comment}  
\usepackage{dednat}
\usepackage{color}

\newcommand{\wrong}{\textsc{\small wrong-}}

\begin{document}
%
%\frontmatter          % for the preliminaries
%
\pagestyle{headings}  % switches on printing of running heads

\title{Gentzen-Prawitz Natural Deduction \\ as a Teaching Tool}
%
% abbreviated title (for running head)
\titlerunning{Prawitz's Style Natural Deduction as a Teaching Tool}  
\author{Jean-Fran\c{c}ois Monin \and Cristian Ene \and Micha\"el P\'erin}
\authorrunning{Jean-Fran\c{c}ois Monin}   % abbreviated author list (for running head)
%
%%%% modified list of authors for the TOC (add the affiliations)
%\tocauthor{Jean-Fran\c{c}ois Monin (Universit\'e de Grenoble)}
%
\institute{Universit\'e de Grenoble 1 %\\
%\email{jean-francois.monin@imag.fr}
}

\maketitle              % typeset the title of the contribution

\begin{abstract}
We report a four-years experiment in teaching reasoning 
to undergraduate students, ranging from weak to gifted,
using Gentzen-Prawitz's style natural deduction.
We argue that this pedagogical approach
is a good alternative to the use of Boolean algebra for 
teaching reasoning,
especially for computer scientists
and formal methods practioners.
% We advocate
\end{abstract}

\section{Introduction}

Logic is one of the uppermost basic ingredients of formal methods.
The most common approach for teaching logic 
takes its root in the model theoretical view:
logical connectors are seen as Boolean functions
(truth tables),
and then generalized to quantifiers:
$\forall$ is like an infinite conjunction,
$\exists$ is like an infinite disjunction.
Teaching logic along these lines
is a well-established tradition.
Boolean algebra proved efficient for solving enigmas and, much more seriously,
for designing digital circuit or automatizing the resolution
of large combinatory problems (e.g.\ by reduction to SAT).
In the area of hardware and of programming, 
they provide Boolean expressions and play a key role
in control structures such as the \textsf{if} and \textsf{while} constructs,
not to speak about bit-level programming.

However, the Boolean approach is not so clearly related to 
\emph{usual} reasoning.
The case of implication is especially questionnable.
In every day life, as well as in mathematics textbooks,
nobody proves an implication $A\impl B$ by computing a truth table:
one assumes $A$ and then proves $B$ under this hypothesis.

It is even argued that logic is essentially about \emph{proofs},
before being about \emph{truth}.
First we can observe that in some logics, including
temporal logics and modal logics which have many applications
in formal methods, the semantics of a proposition is rather more
complex than an truth value --
typically, it is described by a Kripke semantics.
But even in the case of usual logics,
logicians following Dummet, Prawitz and Schroeder-Heister worked 
on a proof-theoretic semantics of logic
(see \cite{schr:vali06} for a recent presentation).

Proofs can be formalized using syntactic objects described by
\emph{deduction systems}.
Such systems were introduced in the last century in order to 
study the meta-theory of logic.
Four years ago, we decided to experiment the use of a particular deduction system,
namely \emph{Gentzen-Prawitz Natural Deduction}
(GPND, for short) for teaching purposes at an introductory undergraduate level.
The choice of GPND is discussed below in section~\ref{sec:nd}.

The main thesis of this paper is that GPND has
a strong pedagogical interest independent from meta-theoretical considerations.
First, it provides a much better explanation of the meaning (and even: essence)
of logical connectors and quantifiers. 
A formal framework for proof writing is necessary to point out
their mistakes in reasoning which, due to ambiguity,
are always arguable in proofs written in natural language.
Manual proof checking becomes perfectly rigourous
-- indeed it can be automatized, but this is another issue --
and moreover
it provides precious hints for proof search.
More technical advantages are discussed below in section~\ref{sec:pedlog}.
Though this material is especially
relevant as an introduction to formal methods,
we claim that, more generally,
it illustrates several key notions of computer science:
\begin{itemize}
\item Case analysis
\item Tree-like data structures
\item Modularity
\item Divide and conquer problem solving
\item Variables and scopes
\item Rule-based formalisms (preparation to more advanced courses)
\item Good support for discussing the relation between syntax and semantics
\item Introduction to proof-assistants
\end{itemize}
Let us add that it also provides a good help for writing rigourous and accurate
proofs by induction.
However, some pitfalls have to be avoided.
The way some notions are introduced is sensitive,
and some notations have to be carefully designed
in order to keep manageable size of interesting proofs
without loss of precision.

Our thesis is supported by our experience with the use of GPND in an introductory
course on logic, given to first year undergraduate students from 2005 to 2009.
The rest of the paper is organized as follows.
In Section \ref{sec:nd}, we present the scientific background,
i.e.\ a short account of natural deduction.
In section \ref{sec:cours}, we outline the contents of the course we gave since 2005.
In section \ref{sec:issues}, we discuss some issues related to the previous experiment
as well as possible extensions.
We conclude in section \ref{sec:conl}.

%%%%%%%%%%%%%%%%%%%%%%%%%%%%%%%%%%%%%%%%%%%%%%%%%%%%%%%%%%%%%%%%%%%%%%%%%%
\section{Background: Natural Deduction}
\label{sec:nd}

Natural Deduction was invented by Gerhard Gentzen \cite{Gen35}
and further studied by Dag Prawitz \cite{Prawitz-ND-65}
for the meta-theoretical study of first-order logic.
In contrast with Hilbert's style deduction systems,
characterized by few inference rules and many axioms,
Gentzen's systems have only one axiom and many inference rules.
A strong point of his approach is that each connector
is considered separately, providing 
a intrisic meaning for it: intuitively,
each connector $*$ is defined by the canonical way
to prove a formula having $*$ as its principal connector
(introduction rules),
or to exploit a formula having $*$ as its principal connector
(elimination rules).
The rules are recalled in figure \ref{fig:GPND}.
All hypotheses have a name such as $h_n$.
Discharged hypotheses are distinguished by square brackets
around their name (e.g. $[h_n]$),
and the place where they are discharged appears in the label
of the rule (then we know that this hypothesis is no longer
available below this rule).

%%%%%%%%%%%%%%%%%%%%%%%%%%%%%%%%%%%%%%%%%%%%%%%%%%%%%%%%
\begin{figure}

% CONJUNCTION

\hfill
\begin{minipage}[t]{6em}
\AxiomC{$A$}
\AxiomC{$B$}
\etintro{A\et B}
\DisplayProof
\end{minipage}
\hfill
\begin{minipage}[t]{5em}
\AxiomC{$A\et B$}
\etelimg{A}
\DisplayProof
\end{minipage}
\hfill
\begin{minipage}[t]{5em}
\AxiomC{$A\et B$}
\etelimd{B}
\DisplayProof
\end{minipage}
\hfill\mbox{}

% IMPLICATION
\bigskip

\hfill
\begin{minipage}[b]{6.3em}
\AxiomC{\begin{tabular}{c}\lhyp{A}{h_n}\\\vdots\\$B$\end{tabular}}
\implintro{A \impl B}{h_n}
\DisplayProof
\end{minipage}
\hfill
\begin{minipage}[b]{8.4em}
\AxiomC{$A\impl B$}
\AxiomC{$A$}
\implelim{B}
\DisplayProof
\end{minipage}
\hfill\mbox{}

% DISJUNCTION
\bigskip

\hfill
\begin{minipage}[b]{5em}
\AxiomC{$A$}
\ouintrog{A \ou B}
\DisplayProof
\end{minipage}
\hfill
\begin{minipage}[b]{5em} 
\AxiomC{$B$}
\ouintrod{A \ou B}
\DisplayProof
\end{minipage}
\hfill
\begin{minipage}[b]{15em} 
\AxiomC{\begin{tabular}{@{}c@{}}\phantom{\lhyp{A}{h_n}}\\\phantom{\vdots}\\$A \ou B$\end{tabular}}
\AxiomC{\begin{tabular}{@{}c@{}}\lhyp{A}{h_n}\\\vdots\\$C$\end{tabular}}
\AxiomC{\begin{tabular}{@{}c@{}}\lhyp{B}{h_m}\\\vdots\\$C$\end{tabular}}
\ouelim{ C}{h_n}{h_m}
\DisplayProof
\end{minipage}
\hfill\mbox{}

% ABSURD, NEGATION
\bigskip

\hfill
\begin{minipage}[t]{3em}
\AxiomC{$\absu$}
\botelim{A}
\DisplayProof
\end{minipage}
\hfill
\begin{minipage}[t]{8em}
$\neg A ~\egdef~ A\impl \absu$\\\mbox{}
\end{minipage}
\hfill\mbox{}
%
\begin{comment}
\begin{minipage}[t]{6em}
\AxiomC{$\neg A$}
\regledef{\neg}{A\impl \absu}
\DisplayProof
\end{minipage}
%
\hfill
\begin{minipage}[t]{6em}
\AxiomC{$A\impl \absu$}
\regledef{\neg}{\neg A}
\DisplayProof
\end{minipage}
\end{comment}
%

% CLASSICAL LOGIC
\bigskip

\hfill
\begin{minipage}[t]{6em}
%\AxiomC{}
\tiersex{A \ou \neg A}
\DisplayProof
\end{minipage}
\hfill
\begin{minipage}[t]{6em}
\AxiomC{$\neg\neg A$}
\nnelim{A}
\DisplayProof
\end{minipage}
\hfill\mbox{}

% QUANTIFIERS FORALL
\bigskip

\hfill
\makebox{
\begin{minipage}[t]{9.5em}
\AxiomC{\begin{tabular}{c}
          \hyp{{H_1}(\_)}{h_1}\ldots
          \hyp{{H_n}(\_)}{h_n}\\
         \vdots\\$P(x_0)$\end{tabular}}
\qqsintro{\qqs{x}P(x)}
\DisplayProof
\end{minipage}
}
\hfill
\begin{minipage}[t]{8em} %\mbox{}\\[3cm]
\vspace*{0em}
%\framebox{
\hspace*{0em}\begin{minipage}[b]{7em}
\AxiomC{\begin{tabular}{c}$\qqs{x}P(x)$\end{tabular}}
\qqselim{P(t)}{x}{t}
\DisplayProof
\end{minipage}
%}
\end{minipage}
\hfill\mbox{}

\medskip
Side conditions for  ${\forall \mathrm{I}}$ :
$x_0$ must not be free in any available hypothesis $h_1$\ldots $h_n$.

% QUANTIFIERS EXISTS
\bigskip

\hfil
\begin{minipage}[b]{8em} %\mbox{}\\[3cm]
\vspace*{0em}
\makebox{
\hspace*{0em}\begin{minipage}[b]{5em}
\AxiomC{\begin{tabular}{c}$P(t)$\end{tabular}}
\exintro{\exi{x}P(x)}
\DisplayProof
\end{minipage}
}
\end{minipage}
\hfil\makebox{
\hspace*{0em}\begin{minipage}[b]{11em}
\AxiomC{\begin{tabular}{@{}c@{}}
         \phantom{\lhyp{{P}(x_0)}{h_n}}\\\phantom{\vdots}\\$\exi{x}P(x)$
	\end{tabular}}
\AxiomC{\begin{tabular}{@{}c@{}}
         \lhyp{{P}(x_0)}{h_n}\\\vdots\\$C$
         \end{tabular}}
\exelim{C}{h_n}
\DisplayProof
\end{minipage}
}
\hfil

\medskip
Side conditions for ${\exists_\mathrm{E}}$ :
\begin{itemize}
\item 
in the proof of $C$ from $P(x_0)$, 
$x_0$ must not be free in any available hypothesis but $h_n$; 
\item $x_0$ must not be free in $C$.
\end{itemize}

\caption{Gentzen-Prawitz Natural Deduction Rules}
\label{fig:GPND}
\end{figure}
The main meta-theoretical property of Gentzen's systems
is the cut-elimination theorem saying that,
basically, proofs can be normalized is a way such that the last rule is
an introduction rule for the principal connector of the conclusion 
\cite{Prawitz-ND-65,Giraflor}.
Important corollaries are the subformula property
(delimiting the proof search space)
and the consistency of logic 
(without reference to model-theoretic semantics).

Here, we are more interested in the pedagogical value
of GPND proof-trees.
We think that it comes from their intrisic features:
they are concrete, intuitive, and
according to some logician philosophers,
they reflect exactly proof objects 
(again, see \cite{schr:vali06}).
The latter fact is particularly evident if one considers
GPND proof-trees as another syntax for typed lambda-calculus,
through the Curry-Howard-De Bruijn isomorphism
\cite{CurryFeys58,Howard69,Debruijn68}
(it is not the case in another popular representation
of natural deduction using sequents, see \ref{sec:seq} for details).

If we compare with the Boolean approach to teaching logic,
proof-trees are certainly more complex than Boolean values but,
to some respect, they look concrete and may be perceived
as less abstract than Boolean functions.
We just may regret that shortcuts using Boolean algebraic laws are not for free
in natural deduction.
In Boolean algebra, equivalence is the same as equality,
whereas here, it is a congruence:
we can show that if $A \eqlo B$, then for any context $\mathcal{C}[\,.\,]$,
we have $\mathcal{C}[A] \eqlo \mathcal{C}[B]$.
The proof is by induction on the structure of contexts.
It is not very difficult and can be understood by good students, 
at least for propositional logic,
and is a good introduction to the metatheorical study of logic
but we could not afford to present it 
at the level considered in our pedagogical experience.

Fortunately, it turns out that algebraic laws are mainly useful
when handling with large propositionnal formulas,
which is not the case in our exercises.
In places where, say, commutativity, associativity
or replacement of $\neg A \impl \neg B$ with $B\impl A$,
could be used, they can easily be bypassed.

%%%%%%%%%%%%%%%%%%%%%%%%%%%%%%%%%%%%%%%%%%%%%%%%%%%%%%%%%%%%%%%%%%%%%%%%%%
\section{Course Outline}
\label{sec:cours}

The course was designed to introduce logical reasoning to students
without previous systematic exposition to logic,
in order to prepare them to further courses on computational
models, automata and languages, program specification and verification, 
formal methods, etc.
Despite some basic practice in mathematics, many of them have
gaps in dealing with proofs and even in capturing the
meaning of implication and quantifiers.

Our aim is then to provide an intuition of logical connectors
and proofs using 
1) a systematic approach based on the structure of
  the formula to prove,
2) a careful and explicit treatment of quantifiers and
3) a computational data-structure able to implement
these requirements, namely proof-trees.

\subsection{Proof Trees}
\label{sec:ptmod}

Just to start with, we assume an intuitive and rough 
knowledge of $\et$, $\ou$ and $\impl$.
The first new idea to become familiar with is the notion of proof-tree.
A difficulty with GPND is that deductions, in general,
depend on hypotheses and that the stock of hypotheses
vary when one progresses in the reading of a proof. 
We then chose to postpone this issue and to use, in the first lesson,
only deductions having no effect on available hypotheses. 
To this effect we could start with GPND rules such that
$\et_{\intr}$, $\et_{\eli[1]}$, $\et_{\eli[2]}$, $\impl_{\eli}$, $\ou\intr[1]$, $\ou\intr[2]$. 
However this would break a systematic exposition of the rules.
We therefore slightly cheat in a first stage: 
we introduce ad-hoc inference rules,
relevant to the formalization of a toy reasoning,
such as \textsc{tri} (transitivity of implication) and \textsc{dli} 
(disjunction on the left of an implication).

\hfil
\begin{minipage}[t]{12em}
\begin{prooftree}
\AxiomC{$A \impl B$}
\AxiomC{$B \impl C$}
\RightLabel{\textsc{tri}}
\BinaryInfC{$A \impl C$}
\end{prooftree}
\end{minipage}
\hfil%\framebox{
\begin{minipage}[t]{12em}
\begin{prooftree}
\AxiomC{$A \impl C$}
\AxiomC{$B \impl C$}
\RightLabel{\textsc{dli}}
\BinaryInfC{$(A \ou B) \impl C$}
\end{prooftree}
\end{minipage} %}

\begin{figure}
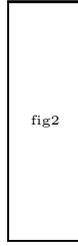

\pgfdeclareimage[height=32mm]{fig2}{branche_arb}
\hfil\pgfuseimage{fig2} 
\caption{Branching proof-trees together}
\label{fig:bran}
\end{figure}

\begin{figure}
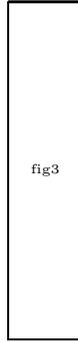

\pgfdeclareimage[height=45mm]{fig3}{imbr_arb}
\hfil \pgfuseimage{fig3} 
\caption{Abstraction of a proof-tree}
\label{fig:abst}
\end{figure}

\medskip\noindent 
The specific rules are not important at this stage.
We aim at teaching a new game, where the key ideas are:
\begin{itemize}
\item The notion of inference rule, with premises, conclusion, 
  and justification (a name used as a label for an inference rule).
\item Checking that an inference rule is correctly applied is an easy
  mechanical task.
\item A rule is actually a schema
  (propositional variables can be replaced with any proposition).
\item A proof-tree relates hypotheses (on the top) to one conclusion
  (at the bottom).
\item A proof-tree is built from inference rules.
\item Generalization and modularity: 
  one can build a proof tree from subtrees.
\end{itemize}
The last items are illustrated in a very intuitive way, using examples 
following the diagrams of figures \ref{fig:bran} and \ref{fig:abst},
where numbers represent propositions\footnote{%
Technically, natural deduction 
distinguishes the name (here: a number) of a proposition and
what the proposition stands, e.g.\ $A\et B \impl C\ou D$.
It may even happen that two different names stand for
the same proposition.
Of course such details are beyond the scope of the lesson.
}.
Figure \ref{fig:abst} also illustrates a situation where the
same hypothesis can be used several times\footnote{%
In general, a hypothesis can be used 0, 1 ore several times.}.
The rules of the game change very quickly (from the second lesson), 
but not its shape.
Actually, playing with somewhat complicated rules
(involving 3 or 4 occurrences of connectors)
drives us to a quest for convincing elementary rules.

Before going further, let us mention that we can \emph{name}
proof-trees and use such names as justifications for
non-elementary proof steps.
We introduce in this way derived inference rules,
in advanced chapters.

%%%%%%%%%%%%%%%%%%%%%%%%%%%%%%%%%%%%%%%%%%%%%%%%%%%%%%%% 

%%%

%%%%%%%%%%%%%%%%%%%%%%%%%%%%%%%%%%%%%%%%%%%%%%%%%%%%%%%%

\subsection{Propositional Connectors}
\label{sec:prop}

We keep the presentation of Gentzen, 
where each rule deals with only one connector.
In some sense, GPND rules provide a semantics to the corresponding connector,
by explaining the canonical ways to prove and to exploit a formula governed by
this connector.
We start with $\et$, which has the simplest rules,
and illustrate them on a proof of $B\et A$ assuming $A\et B$.

The next connector ($\impl$) is the most important for several reasons:
\begin{itemize}
\item 
  any theorem has at least one occurence of $\impl$
  (unless PEM -- Principle of Excluded Middle -- is used) :
  a theorem is the conclusion of a proof-tree where all hypotheses
  are discharged;
\item it is often misunderstood by students, and
\item it is the fundamental place for discussing hypotheses management.
\end{itemize}
That said, $\impl\eli$ is well-known 
(just another name for \emph{modus-ponens})
the statement of $\impl\intr$ is very natural, 
as it sticks to the common practice for proving $A \impl B$:
suppose $A$, then prove $B$.

The hypothesis $h_n$ mentionned in ${\impl\intr}_{[h_n]}$
is said to be \emph{available} in the sub-proof-tree concluding to $B$.
This rule has a special interest for computer scientists,
as it illustrates the notion of scope,
which is here applied on hypotheses :
the scope of an hypothesis is its availibility domain.
Note that the scope is here at the level of proof-trees.
In particular, we insist on maintaining a clear separation, using boxes,
between different sub-proof-trees equipped with their hypotheses,
to represent scopes.
%Another application of this notion, applied to
%first-order variables (then at the level of formulas), 
%appears later.

Interesting exercises, ranging from easy to difficult, 
can be proposed using only $\et$ and $\impl$,
or even just $\impl$.
Here are some examples -- note that $\impl$ associates to the right:
\begin{itemize}
\item $A \et B \impl B \et A$ (very easy);
\item $[(A \et B) \impl C] \impl (A \impl B \impl C)$, and conversely:
  intuitively, there are two equivalent ways to express
  the idea of ``and if'';
\item $(A \impl B\impl C) \;\impl\; ((A \impl B) \impl (A \impl C))$;
\item $A \impl B \impl A$ (somewhat troubling);
\item $(A \equi (A \impl B)) \impl B$.
\end{itemize}
In the last one, $P\equi Q$ is an abbreviation for 
$(P\impl Q) \et (Q\impl P)$.
It is basically the essence of diagonal arguments:
replacing $B$ with absurdity $\absu$ 
and the definition of $\neg$ (see below), it means that $A$ 
cannot be equivalent to $\neg A$. 
Here, the argument is developed constructively, 
without case analysis on $A \ou \neg A$.
An interesting challenge is to find a solution without
repeating sub-proof-trees: for more
advanced students, it illustrates the notion of \emph{cut},
similar to a lemma in informal practice.

We finish this part with disjunction.
The two introduction rules are obvious.
The elimination rules corresponds to case analysis
and requires hypotheses management -- hence another
opportunity to discuss on scopes.
Note that students are tempted to invent 
a $\ou\eli$ rule with 2 conclusions,
something like 
\hfil
\begin{prooftree}
\AxiomC{$A \ou B$}
\RightLabel{\wrong$\ou\eli$}
\UnaryInfC{$A \qquad B$}
\end{prooftree}

\noindent
Here we have to explain that a proof-tree with 2
(and then, in general, many) conclusions
is a complicated beast. 
In some sense those conclusions should be handled separately
(otherwise, we could deduce $A\et B$ from the previous deduction).
However, separate deductions starting from $A$ and from $B$
need eventually to be synchronized, once the same
conclusion is reached. But further complications will happen,
typically if $A$ (or $B$) has to be used several times.
So we keep things simple, sticking to the shape of a tree.

Let us mention here another challenging exercise,
which requires a good understanding of $\impl\intr$
\begin{itemize}
\item $((A \ou A \impl C)\impl C) \impl C$
\end{itemize}
It can related to the elimination of double negations,
to come later.

\subsection{Quantifiers}
\label{sec:quant}

Before explaining the rules,
a number of notions are needed on what is usually
called a first-order language. 
Hoowever, a perfectly formal and rigourous presentation 
would be counter-productive at this level.
Students have an operationnel knowledge of 
terms made of function symbols, constants and variables.
So we insist only on sensitive concepts:
free and bound variables, 
their scope (at the level of the syntax of formulas, here),
substitution of free variables. 
Formulas which differ only on the name of free variables
are considered identical.
The rule for $\forall$ elimination is obvious.
We choose to provide the substitution in the label:
$\forall\eli{(\subs{x}{t})}$ means that the 
(free occurences of) variable $x$
will be substituted with $t$.
This level of precision is useful (even needed!) for students,
we go back to this issue in section \ref{sec:issues}. 

Rule $\forall\intr$ raises the sensitive question of fresh variables.
In the premise $P(x_0)$, $x_0$ stands for an \emph{arbitrary} variable.
What does it mean? 
It is easy to explain that ``arbitrary'' means
``not subject to an hypothesis'' or,
more accurately, ``not subject to an available hypothesis'',
hence ``not free in any available hypothesis''.
We could say as well that 
the premise $P(x_0)$ must not be in the scope 
(at the level of proofs)
of an hypothesis on $x_0$. 
Mastering hypotheses handling is then crucial at this stage.
We require students to write explicitly this side condition
as s $x_0 \not\in FV(h_1,\ldots h_n)$ and to check it
during the proof process.

The explanations about $\exists\eli{[h_n]}$ are along the same lines.
We stress that $\exists$ behaves like an infinite version of $\ou$.
Hence it is not surprising that the structure 
of $\exists\eli$ is similar to $\ou\eli$.
But similarly as well, students tend to formalize 
``we know that $\exists x P(x)$; let $x_0$ be the witness such that $P(x_0)$''
by 
\hfil
\begin{prooftree}
\AxiomC{$\exists x P(x)$}
\RightLabel{\wrong$\exists\eli$}
\UnaryInfC{$P(x_0)$}
\end{prooftree}

\noindent
Such a rule leads very quickly to undesired consequences, 
as it behaves as a $\forall\eli{(\subs{x}{x_0})}$.
Indeed, it yields 
\hfil
\begin{prooftree}
\AxiomC{$\exists x P(x)$}
\RightLabel{\wrong$\exists\eli$}
\UnaryInfC{$P(x_0)$}
\RightLabel{$\qqs\intr$}
\UnaryInfC{$\qqs{x}P(x)$}
\end{prooftree}

\noindent 
hence $(\exists x P(x)) \impl (\qqs{x}P(x))$.
This usual error is an opportunity to discuss on the effect
and the consequences of \wrong$\exists\eli$.
%
%More explanations are needed here.
The right $\exists\eli$ rule takes a situation where
one has a proof of $\exi{x} P(x)$ and,
from \emph{any witness} $x_0$ such that $P(x_0)$, 
one can build a proof-tree $\nabla$ having some conclusion $C$,
possibly using other premises. Then, one can infer $C$.
Note that $P(x_0)$ may be used several times in $\nabla$,
while it is not feasible in \wrong$\exists\eli$.
Students agree that each use of $\exists\eli$ would produce a different $x_i$.
Another important intuition is that, in $\forall\eli{(\subs{x}{x_0})}$,
$x_0$ does not come from the premise $\qqs{x}P(x)$
it is typically given by a $\qqs\intr$
below in the proof-tree -- 
the proof is often built in a bottom-up manner, initially.
This is to be contrasted with in $\exists\eli$,
where $x_0$ is a witness contained
in the proof of the premise $\exists x P(x)$.

At this point students are able to find proof-trees
for formulas such as 
$(\exists{x}\qqs{y} R(x, y)) \impl (\qqs{y}\exists{x} R(x, y))$
and to become aware that the converse is not a theorem.

\subsection{Absurd and Negation}

The absurd proposition ($\absu$) has no introduction rule.
We mention that $\absu$ cannot be proved in the empty context,
as a corollary of the cut-elimination theorem
(the latter is only stated, its proof is beyond the scope
of our course).

Negation is defined by $\neg A \egdef A \impl \absu$. 

It is interesting to note that $\absu$ and $\neg$
can be introduced very late,
after first-order notions. 
Many interesting exercises can be done
without $\neg$.   
In fact, we could delay further these connectors,
after equalities and induction.
A large amount of logic material can be developed without reference to False.
For instance, it is the case for algebraic properties of $+$ and $\times$
on natural numbers.

However we decided to talk about $\neg$ at this stage
in order to introduce classical reasoning,
using either the Principle of Excluded Middle (PEM)
or $\neg\neg\eli$.
This is also the place for discussing about
constructive reasoning -- something which is 
certainly more sensitive in the framework of computer science
than mathematics.

Among the exercises which can be proposed at this stage,
let us mention some puzzles such as
\begin{itemize}
\item $\neg\neg A \et \neg\neg B \eqlo \neg\neg(A \et B)$
\item $\neg\neg(\neg\neg A \ou \neg\neg B \eqlo \neg\neg(A \ou B))$.
\end{itemize}
They can be proved with $\neg\neg\eli$, but finding a solution
without this rule and without PEM) requires a good understanding
of implication.

% \subsection{Equational reasoning}

\subsection{Equational reasonning}
\label{sec:equa}

If $o_1$ and $o_2$ are ``identical'', 
it is clear that everything we prove about $o_1$, holds for
$o_2$ as well. 
This common kind of equational reasoning, often called Leibniz's law,
is embodied in our framework as equality elimination.
We recall it in Figure \ref{fig:eqrules}, together with 
equality introduction, which is the only general axiom
about $=$, that is, equality is reflexive.
It is easy to derive symmetry and transitivity of $=$
from $=\intr$ and $=\eli$.
In principle, it is possible to present any equationnal
reasoning as proof-trees using only $=\intr$ and $=\eli$.
However, it turns out quite tedious and lengthy.
Therefore we prefer to present an equational reasoning
as illustrated on the right hand of Figure \ref{fig:equational}.
It can be shown (by induction on number of rewriting steps)
that such a proof can be put in the proof-tree format.
Implementing that transformation on proof-trees
could be a programming exercise in a companion course
on functional programming.
This proof $\mathcal{E}_i$ is then abstracted under
the form of a multiple inference step
$\AxiomC{$\hyp{}{h_1}\ldots \hyp{}{h_n}$}
\doubleLine
\RightLabel{$\mathcal{E}_i$}
\UnaryInfC{$ U = Z$}
\DisplayProof
$.
Note that justifications used in $\mathcal{E}_i$ can themselves
refer to separate proof-trees.

\begin{figure}
\hfil
\makebox{
\begin{minipage}[t]{9.5em}
\vspace*{1ex}
\eqintro{t}
\DisplayProof
\end{minipage}
}
\hfil
\begin{minipage}[t]{8em} %\mbox{}\\[3cm]
\vspace*{0em}
%\framebox{
\hspace*{0em}\begin{minipage}[b]{7em}
\AxiomC{$a = b$}
\AxiomC{$P(a)$}
\eqelim{P(b)}
\DisplayProof
\end{minipage}
%}
\end{minipage}

\caption{Rules for equality}
\label{fig:eqrules}

\end{figure}

\begin{figure}

%\bigskip
%\hfill
\makebox{
\begin{minipage}[t]{10.5em}
\vspace*{1ex}
\AxiomC{$\hyp{}{h_1}\ldots \hyp{}{h_n}$}
\doubleLine
\RightLabel{$\mathcal{E}_i$}
\UnaryInfC{$  U=  Z$}
\DisplayProof\vspace{0.5cm}

where $\mathcal{E}_i$ is 
\end{minipage}
}
%
%\hfill
\begin{minipage}[t]{8em} %\mbox{}\\[3cm]
\vspace*{0em}
%\framebox{
\hspace*{0em}\begin{minipage}[b]{4em}
\[
\mathcal{E}_i \left\{
\begin{array}{rl}
  & U \\ 
= & \comvg{justification that $U = V$ provided $h_1\ldots h_n$} \\
  &   V  \\
 \vdots \\
  &   Y \\
= & \comvg{justification that $Y = Z$ provided $h_1\ldots h_n$} \\
 &   Z
\end{array}\right.
\]
\end{minipage}
%}
\end{minipage}
\hfill\mbox{}

%\medskip
%\hfil Introducing equational reasoning in our proof trees.
\caption{Equational reasoning}
\label{fig:equational}
\end{figure}

%%% Local Variables: 
%%% mode: latex
%%% TeX-master: tfminf122b
%%% End: 

\subsection{Definitions}

A very useful device for keeping proofs manageable
is to use \emph{definitions}. 
For instance a sub-formula such as $\qqs{x} P(x) \impl Q(x, y)$
can be abbreviated as $R(y)$,
provided we define $R(u) \egdef \qqs{x} P(x) \impl Q(x, u)$.
Free variables have to be properly taken into account.
The definiendum can be freely replaced with the
corresponding definiens and conversely.
Technically speeking, 
in proof-theory, such steps are not
considered as deductions, but follow from the conversion rule,
which means that 
a proof-tree having $A$ as its conclusion \emph{is}
a proof-tree having $B$ as its conclusion,
when $A \egdef B$ or when $ B\egdef A$
(for a more general setting, we refer the reader 
to \emph{deduction modulo} as defined in \cite{Dowek98theoremproving}).
However, for pedagogical purposes,
we prefer to make such steps explicit,
at least at the beginning.
In order to distinguish such steps from regular deduction steps,
we present them using dot lines instead of plain lines. 
In the previous example, we could then write
\hfil
\begin{prooftree}
\AxiomC{$D \et (\qqs{x} P(x) \impl Q(x, y)) \ou E$}
\regledef{R}{D \et R(y) \ou E}
\end{prooftree}
The reverse replacement can also be done by invoking 
$R\,\mathrm{def}$.
%Abbreviated rules: cf. modular constructs introduced in \ref{sec:ptmod}

\subsection{Set theoretic constructs}

Definitions are extensively used when dealing with
set-theoretic constructs.
Besides the extensionnality axiom
\begin{itemize}
\item $A = B \equim (\qqs{x,} x\in A \equim x \in B)$
\end{itemize}
we have the following definitions:
\begin{itemize}
\item $A \inclus B \egdef (\qqs x x\in A \implm x \in B)$
\item $x \in A \inter B \egdef x\in A \,\etm\, x \in B$
\item $x \in A \union B \egdef x\in A \,\oum\, x \in B$
\item $x \in \emptyset \egdef \absu$
\item $x \in \{a\} \egdef x = a\quad$
\item $x \in \{a_1, \ldots a_n\} \egdef x = a_1 \,\oum\, \ldots\,\oum\, x = a_n$
\item $A \in \mathcal{P}(B) \egdef A \,\inclus\, B$
\end{itemize}
From these definitions we prove convenient derived rules, 
given in Figure \ref{fig:setrules}.
From an epistemological point of view, students get a taste
on mathematical foundations:
relying on a solid basis to define new objects and
some convenient abstract rules to reason about.

\begin{figure}

\hfill
\begin{minipage}[b]{8em}
\begin{prooftree}
\AxiomC{\begin{tabular}{c}\lhyp{x\in A}{n}\\\vdots\\$x\in B$ \end{tabular}}
\qqsintroinclus{A\inclus B}{n}
\end{prooftree}
%\DisplayProof
\end{minipage}
\hfill
\begin{minipage}[b]{12em}
\begin{prooftree}
\AxiomC{$x\in A$}
\AxiomC{$A \inclus B$}
\inclelim{x\in B}
\end{prooftree}
\end{minipage}
\hfill
\begin{minipage}[b]{13em}
\begin{prooftree}
\AxiomC{\begin{tabular}{c}\lhyp{x\in A}{n}\\\vdots\\$x\in B$ \end{tabular}}
\AxiomC{\begin{tabular}{c}\lhyp{x\in B}{m}\\\vdots\\$x\in A$ \end{tabular}}
\racext{A=B}{n}{m}
\end{prooftree}
\end{minipage}
\hfill\mbox{}

\medskip
\medskip

\hfill
\begin{minipage}[b]{10em}
\AxiomC{$x\in A$}
\AxiomC{$x\in B$}
\interintro{x\in A\inter B}
\DisplayProof
\end{minipage}
\hfill
\begin{minipage}[b]{7em}
\AxiomC{$x\in A\inter B$}
\interelimg{x\in A}
\DisplayProof
\end{minipage}
\hfill
\begin{minipage}[b]{7em}
\AxiomC{$x\in A\inter B$}
\interelimd{x\in B}
\DisplayProof
\end{minipage}
\hfill\mbox{}

\medskip
\medskip

\hfill
\begin{minipage}[b]{7em}
\AxiomC{$x\in A$}
\unionintrog{x\in A\union B}
\DisplayProof
\end{minipage}
\hfill
\begin{minipage}[b]{7em}
\AxiomC{$x\in B$}
\unionintrod{x\in A\union B}
\DisplayProof
\end{minipage}
\hfill
%\framebox{
\begin{minipage}[b]{18em}
\AxiomC{\begin{tabular}{@{}c@{}}\phantom{\lhyp{A}{n}}\\\phantom{\vdots}\\$x\in A\union B$\end{tabular}}
\AxiomC{\begin{tabular}{@{}c@{}}\lhyp{x\in A}{n}\\\vdots\\$P$\end{tabular}}
\AxiomC{\begin{tabular}{@{}c@{}}\lhyp{x\in B}{m}\\\vdots\\$P$\end{tabular}}
\unionelim{P}{n}{m}
\DisplayProof
\end{minipage}
\hfill\mbox{}

\caption{Derived rules for set-theoretic notations}
\label{fig:setrules}
\end{figure}

\subsection{Induction}

The usual induction principle is formalized by the following 
deduction rule:

\hfil
\begin{prooftree}
\AxiomC{$P(0)$}
\AxiomC{$\qqs{n}\, P(n)\impl  P(S(n))$}
\natelim{\qqs{n} P(n)}
\end{prooftree}
%\caption{Rule for Induction}
%\label{fig:ind}

%Induction on quantified formulas.

We also provide Peano axioms and then can propose exercises
on elementary algebraic properties of addition and multiplication.

\hfil
\begin{minipage}[t]{8em}
\begin{prooftree}
\plusZ{n}
\end{prooftree}
\end{minipage}
\hfil
\begin{minipage}[t]{15em}
\begin{prooftree}
\plusS{n}{m}
\end{prooftree}
\end{minipage}

\hfil
\begin{minipage}[t]{8em}
\begin{prooftree}
\multZ{n}
\end{prooftree}
\end{minipage}
\hfil
\begin{minipage}[t]{15em}
\begin{prooftree}
\multS{n}{m}
\end{prooftree}
\end{minipage}

\noindent
Predicates $\leq$ and $<$ can be defined by
$m \leq n \egdef \exi k n = k + m$
and $m < n \egdef S(m) \leq n $.
Then we can derive ``strong'' (or noetherian) induction
on natural numbers:

\hfil
\begin{minipage}[t]{15em}
\begin{prooftree}
\AxiomC{$\qqs n (\qqs m m < n \impl P(m)) \impl P(n)$}
\doubleLine
\RightLabel{$_{\mathrm{strong}~ \mathrm{nat}-\mathrm{rec}}$}
\UnaryInfC{$\qqs n P(n)$}
\end{prooftree}
\end{minipage}

A natural extension is to work on structural induction on 
ML-style lists or binary trees.

\subsection{Example}

In order to illustrate the above ideas, we consider the example
depicted in Figure \ref{fig:ancestor}, which is a typical
examination problem. In order to simplify notations,
we do not type variables, but we want to emphasize that $n$ is a
natural variable (hence we can apply inductive reasoning over
it). Also, for sake of simplicity, we start by defining some
predicates. Hence, predicate $H_1$ states that everybody has a
father. Predicates $H_2$ states that everybody is its own
$0$-ancestor, and predicate $H_3$ states that the $n+1$-ancestor is
defined as the father of the $n$-ancestor.  We want to prove that for
any $n\in \nat$, everybody has an $n$-ancestor, that is, for any $n\in
\nat$, the property $Q(n)$ holds. The way we teach this example is
the following one. First, we remark that we want to prove some
statement, without any additional hypothesis. Fortunately, the goal
corresponds to an implication. Hence, it suffices to prove the right
part, admitting the left part as an hypothesis. Formally, this
corresponds to an aplication of an $\impl_I$ rule. We continue this
bottom-up proof by applying the rule $\impl_I$ twice again. 
Now, we face a property of the kind $\qqs n ...$ where $n$ is a
variable ranging over naturals, so we can apply induction.
The next goal is split into two other sub-goals.  
Here, we can see the manner we compose proofs,
for example the tree $T_1$ can be proved separately, and then it can
be plugged in the overall proof. But we have to pay attention to the
hypothesis that remain active at the end of the proof $T_1$.  Other
interesting points are the way we unfold definitions, and the use of
equational reasoning.

\begin{figure}
%\begin{minipage}[t]{10em}
{
%\small
%\footnotesize
%\tiny
\scriptsize

\newcommand{\ha}{\qqs x \exi y y=F(x)}
\newcommand{\hb}{\qqs x x=A(0,x)}
\newcommand{\hc}{\qqs n \qqs x F(A(n,x))=A(S(n),x)}
\newcommand{\cc}{\qqs n \qqs x \exi y  y=A(n,x)}
Let us note
$Q(n) \egdef ~\qqs x \exi y  y=A(n,x)$,
$H_1\egdef \ha$, $H_2\egdef \hb$, $H_3\egdef \hc$.

\newcommand{\but}{}

\renewcommand{\but}{\qqs n \qqs m S(n) + m = S(n+m)}

\newcommand{\tree}[3][]{%
  \AxiomC{\ensuremath{#1}}
%  {\dottedLine\doubleLine\dottedLine\doubleLine}
  {\doubleLine\doubleLine}
  \RightLabel{\ensuremath{#3}}
  \UnaryInfC{\ensuremath{#2}}}

\begin{prooftree}
\AxiomC{\lhyp{H_2}{2}}
\regledef{H_2}{\hb}
\qqselim{x_0=A(0,x_0)}{x}{x_0}
\exintro{\exi y  y=A(0,x_0)}
\qqsintro{\qqs x \exi y  y=A(0,x_0)}
\regledef{Q}{Q(0)}
%%%%%
\tree[\lhyp{H_1}{1}\qquad\lhyp{H_3}{3}]{\qqs m  Q(m) \impl Q(S(m))}{T_1}
%%%%%
%%%%%
%\regledef{Q}{\qqs x \exists y  y=A(m_0,x)}
\natelim{\qqs n Q(n)}
\regledef{Q}{\cc}
%%%%
\implintro{H_3 \impl (\cc)}{3}
\implintro{H_2 \impl (H_3 \impl (\cc))}{2}
\implintro{H_1 \impl ((H_2 \impl (H_3 \impl (\cc)))}{1}
\end{prooftree}

where the tree $T_1$ is 

\begin{prooftree}
\AxiomC{\lhyp{Q(m_0)}{hrec}}
\regledef{Q}{\qqs x \exi y  y=A(m_0,x)}
%\qqselim{\exi y  y=A(m_0,x_0)}{}{}
\qqselim{\exi y  y=A(m_0,x_0)}{x}{x_0}
%%%%%%%%%%%%%%%%%%%%%%%%%%%%%%%%%%%%%%%%%%%%%%%%
\AxiomC{\hyp{H_1}{1}}
\regledef{H_1}{\qqs x \exi y y=F(x)}
%\qqselim{\exists y  y=F(y_1)}{}{}
\qqselim{\exi y  y=F(y_1)}{y}{y_1}

\adp[\lhyp{y_1=A(m_0,x_0)}{e_1}, \lhyp{y_2=F(y_1)}{e_2}, \hyp{H_3}{3}]{y_2=A(S(m_0),x_0)}{\mathcal{D}_1}
\exintro{\exi y  y=A(S(m_0),x_0)}

\exelim{\exi y  y=A(S(m_0),x_0)}{e_2}
%%%%%%%%%%%%%%%%%%%%%%%%%%%%%%%%%%%%%%%%%%
\exelim{\exi y  y=A(S(m_0),x_0)}{e_1}
\qqsintro{\qqs x \exi y  y=A(S(m_0),x)}
\regledef{Q}{Q(S(m_0))}
\implintro{Q(m_0) \impl Q(S(m_0))}{hrec}
\qqsintro{\qqs m  Q(m) \impl Q(S(m))}
\end{prooftree}

and $\mathcal{D}_1$ is 
\[
\mathcal{D}_1 \left\{
\begin{array}{rl}
  & y_2 \\ 
= & \comvg{hypothesis $e_2$} \\
  &   F(y_1)  \\
=  & \comvg{hypothesis $e_1$} \\
  &   F(A(m_0,x_0)) \\
= & \comvg{$H_3$ by hypothesis 3, with $\qqs\eli\subs{n}{m_0}$ and $\qqs\eli\subs{x}{x_0}$} \\
 &   A(S(m_0),x_0)
\end{array}\right.
\]
}
%\end{minipage}

\caption{Example ``Everybody has n-ancestors''}
\label{fig:ancestor}
\end{figure}

%%%%%%%%%%%%%%%%%%%%%%%%%%%%%%%%%%%%%%%%%%%%%%%%%%%%%%%%%%%%%%%%%%%%%%%%%%
\section{Pedagogical Issues and Assessment}
\label{sec:issues}

\subsection{Technical Advantages of GPND for Teaching Logic}
\label{sec:pedlog}

We already mentionned that GPND proof-trees reflect usual reasoning
much better than the truth-table approach. 
This is especially clear on implication. 
The fact that 
the rules for $\impl$ and for $\et$ look completely different 
is a chance, as it helps beginners not to confuse between these two connectors.

A very interesting point, from a pedagogical perspective,
is that proof trees allow us to point out errors with accuracy.
It occurs quite often that some student comes with rough and obscure ideas
and still believes that his argument is good. 
It is much more difficult to show him where are his mistakes
on informal or semi-formal writings than on proof trees.
If a rule is wrongly applied, we can first say
``this does not conform to the you Law, on which we agreed'' and
may add:
``if your rule was right, you would get this or that undesired consequence''.
This turns out quite convincing with all kind of students.
Here are some typical errors that can be pointed out in GPND
proof style:
\begin{itemize}
\item Incorrect use of a deduction rule (especially $\ou\eli$).
\item Violation of the scope of an hypothesis
  (for instance, a hypothesis available in a branch of a case analysis
  is used in the other branch).
\item Violation of the side condition of $\exists\eli$ or $\qqs\intr$,
  using a ``convenient'' choice for $\exists x$ or $\qqs x$
  instead of a fresh variable.
\item Exploiting an implication, or a rule, without proving
  its premises.
\end{itemize}
Such errors correspond to typical wrong reasoning written informally.

\medskip
Another benefit is that GPND enforces a precise understanding
of the distinction between available and discharged hypotheses.
This is particularly important for a rigorous treatment
of quantifiers and of inductions,
especially when we they are embedded.
It is sometimes crucial, when proving
a property of the form $\qqs n \qqs m P(n, m)$
by induction on $n$, that the inductive property
$\qqs m P(n, m)$
remains universally quantified,
because the $m$ we need for $S(n)$ may come from a different
value in the induction hypothesis.
A well-known example, among others, is the proof of 
strong induction
using basic induction on the property
$\qqs m m\leq n \impl P(m)$.

\medskip
Let us now mention a number of issues showing the relevance 
of the GPND approach to computer science, including formal methods.

\subsubsection{Case analysis.} 
The elimination rule for $\ou$ states exactly how a disjunctive
piece of information can be exploited.
This is clearly related to algorithmic constructs 
such \textsf{if}\ldots \textsf{then}\ldots \textsf{else}\ldots,
\textsf{case}\ldots \textsf{of}\ldots, 
\textsf{switch}\ldots

\subsubsection{Tree-like data structures.} 
Trees are a ubiquitous concept in computer science.
Their handling is intuitive.
Here is an opportunity to introduce and manipulate them
at an abstract level, 
without reference to an implementation. 

\subsubsection{Modularity.} 
Even middle-size proofs cannot be displayed using
a monolithic proof-tree on a single sheet of paper.
Structuring a proof in sub-trees with a clear interface
allows one to handle this issue.
Here, the interface is defined by the set of hypotheses
and the conclusion. 

\subsubsection{Problem solving.} 
Faced to proving a goal from given hypotheses,
many steps can be carried out just by examining
the form of the formulas at hand.
If the conclusion is among the hypotheses, we are done.
Else, if the conclusion is not atomic, 
it can be decomposed along a divide-and-conquer approach,
using an introduction rule.
However, some of them ($\ou\intr[1]$, $\ou\intr[2]$ and $\exists\intr$)
are dangerous as they may drive into a dead end
We warn the students to postpone as far as possible the use of these rules.
These are the places where thinking on the contents of hypotheses
and creativity are needed.
This method can be carried out in parallel on the corresponding
informal reasoning.

\subsubsection{Variables and scope.} 
It is clear that the notions of free, bound variables 
with their scopes are developed 
when introducing the syntax of first-order formulas.
Admittedly, this comes from formal logic, not specifically GPND.
Looking at programming languages,
logic variables are closer to the concept to be
found in the functionnal paradigm than in the imperative
paradigm. 

What is more specific to GPND is that scopes are also
related to hypotheses, more exactly names of hypotheses:
the scope of an hypothesis is the largest sub-treee where 
it is available (still not discharged).
One can even speak about local and global hypotheses.
Scopes are then discussed already in the framework of propositional
logic.

\subsubsection{Rule-based formalisms.} 
Many formalisms used in computer science a rule-based 
presentation: typing systems, operational semantics, etc.
Studying GPND is then a good training in order to prepare
more advanced courses.

%\subsubsection{Good support for discussing the relation between syntax and semantics.} 

\subsubsection{Introduction to proof assistants.} 
In the area of formal methods and software verification,
well-recognized proof assistants such as Coq \cite{CoqManualV82,BC04}
and Isabelle/HOL \cite{Nipkow-Paulson-Wenzel:2002} 
are available and commonly used.
Their theoretic foundations are logic systems quite close to GPND.

%%%%

\subsection{Pedagogical Issues and pitfalls}
\label{sec:pitfalls}
A number of pedagogical issues have to be taken into account,
in order to make GPND an efficient tool for teaching.

We already mentioned that
proof trees allow us to point out mistakes with accuracy.
To this effect, we insist that 
justifications (labels used in deduction steps) are mandatory: 
often students suffer from a lack of precise ideas
on what their are really doing, or at least
from a poor ability to communicate their arguments.
In this spirit we tend to demand more than
what is generally given in textbooks. 
For instance, the elimination rule for $\forall$
makes explicit the term $t$ to be substituted to
the quantified variable $x$: $\forall\eli{(\subs{x}{t})}$.

%\item Zoom effect

There is a pedagogical issue with implication: 
in GPND, it is impossible\footnote{%
At least for intuitionistic logic.
But it is clear that PEM cannot be exploited without
hypothesis management, 
except if the conclusion is just an instance of $A \ou \neg A$.
For example, all theorems of the form $\neg P \ou Q$,
where $Q$ can be deduced from $P$, are proved by case analysis
on $P \ou \neg P$, i.e.\ using $\ou\eli$.
} to get a theorem without using $\impl\intr$.
Hence we have to consider hypothesis management very early.
As explained in \ref{sec:ptmod}, we fix this issue
by temporary considering fake rules such as
 \textsc{tri}  and \textsc{dli}.
A drawback is that some students tend to think that
any rule is good, provided it looks plausible.
So we have to insist heavily, from the beginning,
that \textsc{tri}  and so on should be forgotten
and that the right rules are coming. 
A complementary approach is to work on deductions
under hypotheses with simple rules such as $\impl\eli$
$\qqs\eli$ and rules for $\et$.
In fact we let students discover these rules in
the first exercise session.

We already mentionned some wrong attempts about
$\ou\eli$ and $\exists\eli$, respectively in
\ref{sec:prop} and \ref{sec:quant}.
How to deal with space consuming proofs
was considered from the beginning (\ref{sec:ptmod}),
through modularity, and using a special notation
for equational reasoning (see \ref{sec:equa}).

%\subsection{Tips and Tricks}
%\label{sec:tricks}

%coups tordus dixit Micha\"el

\subsection{An alternative to GPND: sequents}
\label{sec:seq}
Natural deduction can be presented in terms of \emph{sequents}
$\Gamma\vdash C$, where 
$\Gamma$ is a multiset of formulas and $C$ a formula,
whose intuitive meaning is
``given the conjunction of hypotheses in $\Gamma$,
the conclusion $C$ holds''.
Rules have several sequents as premises and a sequent 
A formula $A$ is a theorem if the sequent $\vdash A$ can be derived.
A \emph{proof tree} is a tree labelled as follows:
leaves are labelled by axioms, i.e.\ sequents $\Gamma\vdash C$ where $C \in \Gamma$.
Although this approach may be prefered
for the meta-theoretical study of natural deduction \cite{gallier93},
it puts the emphasis more on provability than on proofs.

Note that, 
from a pedagogical perspective, the sequent based presentation
is closer to inference systems used 
for typing or structured operational semantics.
But it is a bit far from the objectives of an introductory course to logic.
Moreover, once somebody is familiar with a deduction system,
it is reasonnable to expect that she or he can easily move
to another presentation of it or to another inference system.

\subsection{Assessment} 

This course was given to an audience of
150 to 200 students per year, with the following
rythm: one lecture (1h30) and one exercise class (1h30) per week,
during 11 weeks.
About 3 weeks are needed for discovering proof-trees,
the 3 first propositionnal connectors;
then another 3 weeks for quantifiers and negation;
the next 3 weeks are devoted to set-theoretic notions
ans the last 2 weeks to induction.

The course got a good ranking from the students,
which is quite satisfactory since most of the material
is new for them.
It turns out that they like to play with trees.
However,
our main goal in introducing formal proofs as a first year course was
to improve the ability of students in reasoning beyond the formal
framework of GPND, that is detecting wrong deduction and convincing
proof in natural language.
We brought students from approximate reasoning in natural language
to formal proofs, and we expected them to do the opposite by themselves.
After four years we had to admit that we partially failed. Indeed, some
students were still handicapped when asked to prove a statements in
natural language whereas they were perfectly able to build the proof
tree in GPND style.
This observation brought us to the conclusion that we need more
time transferring the lessons learned in GPND back to the free reasoning.
This includes proof guidelines:
how to decompose a statement to proof into subgoals,
how to find the hypothesis and what should finally be proved
and
how a proof tree can be told as an argumentative discourse.
On this last point we plan to extend the teaching with a project -- in
collaboration with a course on functional programming -- that consists
in a systematic translation of a GPND proof tree into a reasoning in
natural language.

\begin{comment}
This is both a good point and a danger,
because we mainly expect them to become more familiar
with rigorous reasoning. 
They should be able to translate proof-trees into
rigorous explanations.

However, it seems that, for many of them, more time is required 
to become fluent enough with the usual presentation of proofs.
An interesting exercise is to provide
a mechanical recursive procedure which, given a proof-tree,
prints a textual representation of a proof.
Up to now this exercise was carried out informally,
with pencil, paper and drawings.
We expect to organize a formal project based on this idea,
in relation with a neighbour course on functionnal programming.
\end{comment}

We are anyway convinced that working on proof-trees help students to
get a more structured mind.
In order to strengthen the work done so far,
connections have to be established with other courses given
in second year on
logic, automata and languages,
proofs and algorithmics.
Students happen to ask for building proof-trees
2 or 3 years later.
When one of them is stuck at the beginning of a proof,
suggesting her or him to start a proof-tree 
turns out quite helpful.
On the teaching side,
colleagues have to be convinced that
our approach is good and can be reused to some extent.
We are confident that progress will be done in this direction,
because our teaching team became quite enthusiastic,
though most teachers discovered natural deduction
in this course.

\begin{comment}
\paragraph{Limitations and improvements.}

Our main goal in introducing formal proofs as a first year course was
to improve the ability of students in reasoning beyond the formal
framework of GPND, that is detecting wrong deduction and convincing
proof in natural language.
%
We brought students from approximate reasoning in natural language
to formal proofs, and we expected them to do the opposite by themselves.

After four years we had to admit that we partially failed. Indeed, some
students were still handicapped when asked to prove a statements in
natural language whereas they were perfectly able to build the proof
tree in GPND style.
%
This observation brought us to the conclusion that we must spend some
time transferring the lessons learned in GPND back to the free reasoning.
%
This includes proof guidelines:
how to decompose a statement to proof into subgoals,
how to find the hypothesis and what should finally be proved)
and
how a proof tree can be told as an argumentative discourse.
%
On this last point we plan to extend the teaching with a project -- in
collaboration with a course on functional programming -- that consists
in a systematic translation of a GPND proof tree into a reasoning in
natural language.
\end{comment}

%%%%%%%%%%%%%%%%%%%%%%%%%%%%%%%%%%%%%%%%%%%%%%%%%%%%%%%%%%%%%%%%%%%%%%%%%%
\section{Conclusion}
\label{sec:conl}
We advocated that GPND is the good way to introduce logic to beginners,
at least for students in computer science. 
Everybody gets a chance to better understand 
what is a reasoning
and to improve her or his reasoning abilities.
What about other scholars ?
It is often advocated that computer science should be taught much
earlier in the curriculum, notably in the highschool.  Computer
scientists should contribute to this chapter of mathematics.
In particular, proof-trees are simple to understand and funny.
They require no mathematical background.
We think that they could be introduced at the highschool,
at least for propositional logic,
thus helping scholars in their scientific activities. 

Let us finish with some perspectives. 
We limited ourselves to a pencil and paper approach,
mainly because we didn't have enough time slots to do otherwise.
We plan to use a proof assistant in a next version of the course.
However, we will have to take care of the danger
of button-pushing:
existing proof assistants are good at helping the user
to find proofs and to automatize tedious tasks.
Hoawever we want here the user to be aware of the
elementary deduction steps.
In a pedagogical use, a proof-assistant should just
be used as a proof checker.

\bibliographystyle{plain}
\bibliography{biblio}

\end{document}